\documentclass[a4paper,fleqn,usenatbib]{mnras}  
\usepackage{newtxtext,newtxmath}
\usepackage[T1]{fontenc}
\usepackage{ae,aecompl}
\usepackage{graphics,graphicx}
\usepackage{amsmath}	

\title[Abell 2244 to the cluster outskirts]{The flat entropy profile at the outskirts of the Abell 2244 galaxy cluster}
\author[]{S. Andreon$^{1}$, A. Moretti$^{1}$, H. B\"ohringer$^{2,3}$, F. Castagna$^{1,4}$\\ 
$^1$ INAF--Osservatorio Astronomico di Brera, via Brera 28, 20121, Milano, Italy\\
$^2$ Universitats-Sternwarte Munchen, Fakultat fur Physik, Ludwig-Maximilians-Universitat Munchen, Scheinerstr. 1, 81679 München,Germany \\
$^3$ Max-Planck-Institut fur extraterrestrische Physik, 85748 Garching, Germany\\
$^4$ Department of Pure and Applied Sciences, Computer Science Division, Insubria University, 21100 Varese, Italy\\
}
\date{Accepted XXX. Received YYY; in original form ZZZ}
\pubyear{2020}
\begin{document}
\label{firstpage}
\pagerange{\pageref{firstpage}--\pageref{lastpage}}
\maketitle

\begin{abstract}
Entropy is an advantageous diagnostics to study the thermodynamic history of the intracluster plasma of galaxy clusters.
We present the entropy profile of the Abell 2244 galaxy cluster derived both exclusively using X-ray data from the low-background Swift XRT telescope and also using Planck $y$ data. The entropy profile derivation using X-rays only is robust at least to the virial radius because the cluster brightness is large compared to the X-ray background at low energies, temperature is strongly bounded by the lack of cluster X-ray photons at energies $kT>3$ keV, and the XRT background is low, stable and understood. In the
observed solid angle, about one quadrant,
the entropy radial profile deviates from a power-law at the virial radius, mainly because of a sharp drop of the cluster temperature. 
This bending of the entropy profile is confirmed when X-ray spectral information is
replaced by the Compton map. 
Clumping and non-thermal pressure support
are insufficient to restore a power law entropy profile because they are bound to be small by: a) the
agreement between mass estimates from different tracers (gas and galaxies), b) the agreement between entropy profile
determinations based on combinations of observables with different sensitivities and systematics, and c) the low value of clumping as
estimated using the azimuthal scatter and the gas fraction. Based on numerical simulations, 
ion-electron equilibration is also insufficient to restore a linear entropy profile. Therefore, the bending of the entropy profiles seems to be robustly derived and witnesses the teoretically-predicted decrease in the inflow through the virial boundary.

\end{abstract}

\maketitle

\section{Introduction}

Galaxy clusters form by hierarchical accretion of cosmic matter (Peebles 1993). They reach the virial equilibrium over the volume  where the pristine gas accretes onto the dark matter halo through gravitational collapse and is heated up to several million degrees through adiabatic compression and shocks. 
The cluster outskirts are expected to retain most of the information on the processes 
that characterize the accretion and evolution of the main baryon component 
(Roncarelli et al. 2006; Voit 2005).
Entropy is particularly useful in this respect because it is conserved in gravitational processes and it captures
the thermal history of the cluster by recording gains and losses 
of thermal energy.
Therefore, entropy probes the effects of non-gravitational
processes such as supernovae, AGN feedback, and intracluster medium cooling.

In the classical two stages scenario, cluster build up proceeds from inside-out:  the early violent collapse produces
the cluster core followed by mass growth in which the baryons are strongly shocked and entropy is efficiently produced to
give a power-law entropy profile (Zhao et al. 2003, Fakhouri et al. 2011, Wang et al. 2011, Andreon et al. 2021). Eventually, the inflow decreases as the accretion rate peters out and/or become transonic, weakening the virial
shock and therefore entropy production. This process leads to bent entropy profiles near the virial\footnote{We use the expression ``virial radius" to refer to $r_{200}$, the radius at which the interior average density is 200 times the critical density of the Universe.} radius at the present epoch (e.g. Lapi et al. 2010; Cavaliere \& Lapi 2013). The picture is
consistent with numerical simulations (Vazza et al. 2010, Burns et al. 2010, Planelles et al. 2014, Barnes et al. 2017) showing entropy profiles that flatten at the virial radius. 
Indeed, the final entropy profile is a
balance between the hierarchical clustering of the dark matter inducing
gravitational heating of the diffuse baryons pushing the intracluster medium toward a
power-law radial profile and non-gravitational processes pushing the radial profile away from a power-law.
The cluster outskirts therefore are expected to differ
markedly from the cores and present an opportunity to study the cluster gas-dynamics
in a completely different regime.

At large radii, numerical simulations (e.g., Lau, Kratvsov, and Nagai 2009) indicate
that the contribution of non-thermal pressure support increases with radius because
random gas motion becomes increasingly important in the cluster outskirts
due to merging activity, gas accretion, and supersonic motions of galaxies. 
At these radii, gas  clumping  plays  an  important  role (Mathiesen, Evrard, and Mohr 1999; Nagai \& Lau 2011) 
because a clumpy distribution emits more photons than the uniform distribution assumed in the data analysis,
leading to  overestimated densities, gas fraction, and pressure, and an underestimate of entropy.

\begin{figure}
\centerline{\includegraphics[trim=20 0 30 0,clip,width=8.5truecm]{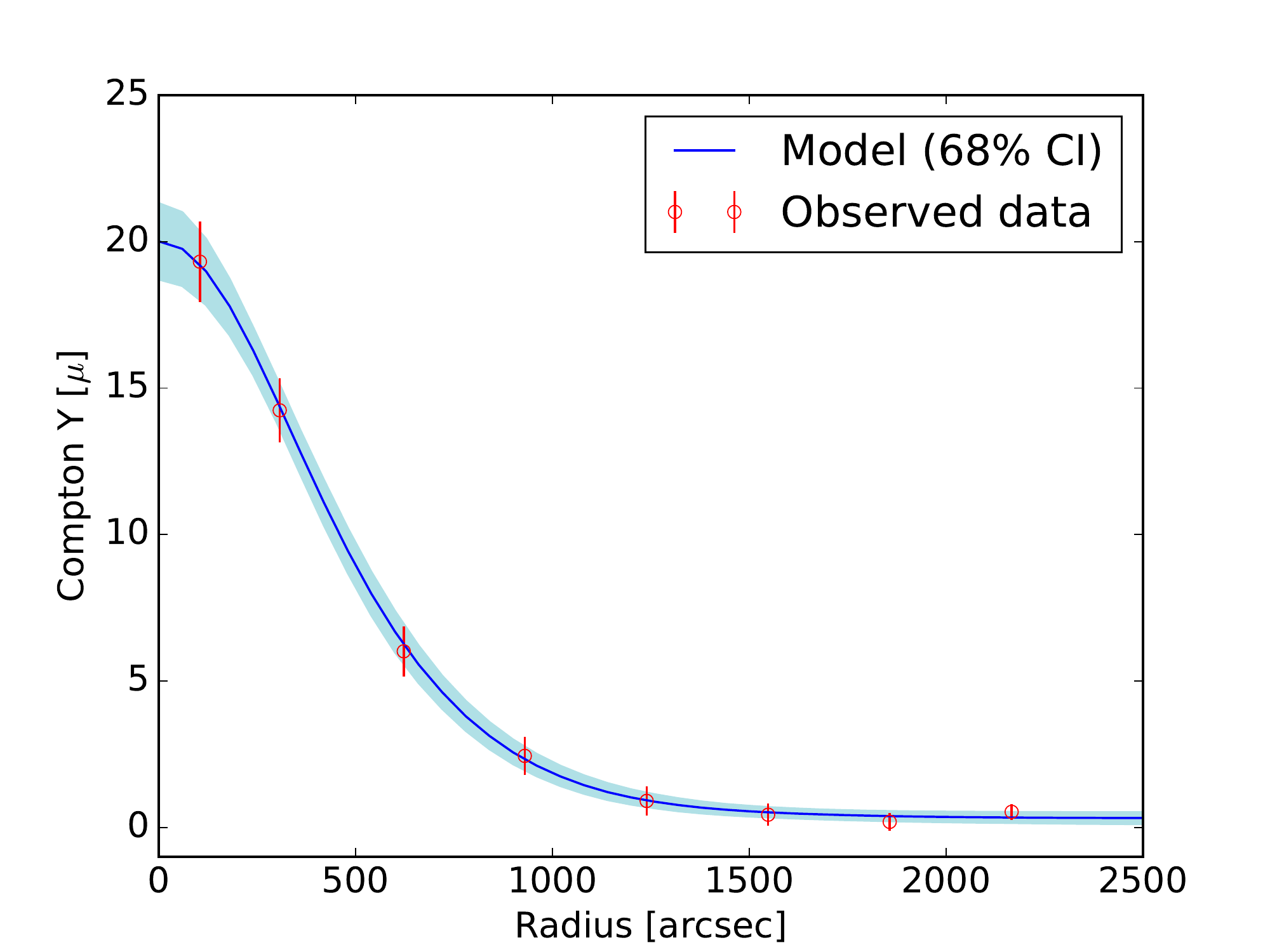}}
\caption[h]{A2244 SZ surface brightness profile (points with error bars) and 68\% uncertainties on the fitted model from Planck data. The adopted spline for the pressure profile fits the observed data well.
}
\label{fig:surf_bright}
\end{figure}

The shape of the entropy profile in the cluster outskirts is still a matter of debate
(see Walker et al. 2019 for a review). 
Several entropy profiles based on data taken with the low-background
Suzaku telescope  claim the existence of
a bending
(e.g., Akamatsu et al. 2011, 2012, Walker et al. 2012,
Ibaraki et al. 2014, Okabe et al. 2014).
Others authors, mostly combining Planck and XMM-Newton, found instead power-law entropy
profiles at large radii (Ghirardini et al. 2019, Mirakhor \& Walker 2020), but
with examples of bent entropy profiles (e.g. Abell 2319 in
Ghirardini et al. 2019). While part of the discrepancy can be explained by the different treatments of clumping, the bent entropy profile after clumping correction of Abell~2319 shows that clumping is not the only reason for the claimed
differences.

The measurements of the properties of the ICM in the cluster outskirts are
performed in a very low Signal/Background regime and are hampered by background systematics.
In the soft band, 0.5-2.0 KeV the main background component is given by local
and Galactic thermal emission (Kunts\& Snowden 2000), while at higher energy it is mostly of
instrumental origin (NXB).
This latter component is particularly annoying in the high-orbit missions,
like Chandra, XMM and eRosita, while the low and stable Suzaku background (Mitsuda et al. 2007) made
outer region obserevation simpler (Reiprich et al. 2009).
However, the Suzaku poor PSF hinder the evaluation of the 
Galactic foreground and cosmic X-ray background contribution (Bautz et al. 2009).

Swift-XRT has an XMM-like PSF (Moretti et al. 2005) and  a stable and well reproducible
instrument background over the whole energy band  (Moretti et al. 2009, 2011, 2012).
This results in the lowest background per unit cluster (or extended source) photon (Walker et al. 2019) among
the all current X-ray telescopes and makes it the best instrument to observe the very low surface brightness sources.
Swift has, however, not been used so far to study cluster outskirts.

In this work, we present Swift observations of the Abell 2244 (A2244) outskirts.
A2244 is at $z=0.0968$ (Zhu et al. 2016). 
A2244 was selected because it is the brightest
clusters with $r_{200}<20$ arcmin, at 
high galactic latitude, and far from the Fermi Bubbles (to have a more easily characterizable background). 
The angular constraint comes from stray-light contamination, affecting radii in the 40 to 60 arcmin range
(Moretti et al. 2009). Based on Chandra data, the cluster has a regular morphology, a flat temperature profile within 500 kpc, and a large value of central entropy
(Donahue et al. 2005).
Rines et al. (2006) measured 
$r_{200}=1.75$ Mpc (corresponding to about 16 arcmin) and $M_{200}= (6.1\pm2.8) \  
10^{14} M_\odot $  using the caustics technique. 

Throughout this paper, we assume $\Omega_M=0.3$, $\Omega_\Lambda=0.7$, 
and $H_0=70$ km s$^{-1}$ Mpc$^{-1}$. 
Results of stochastic computations are given
in the form $x\pm y$, where $x$ and $y$ are 
the posterior mean and standard deviation. The latter also
corresponds to 68\% uncertainties because we only summarize
posteriors close to Gaussian in this way. All logarithms are in base 10.

\section{Data and Analysis}

\subsection{SZ data \& analysis}

To measure the pressure profile from SZ data, we used the data release 2 of the Planck y map (Planck collaboration, 2016) derived using the MILCA component separation algorithms.
We computed the radial profile in annuli of 6 pixels (5.25 arcmin) width (the Planck PSF has $\sigma=4.2$ arcmin).
We adopted the same centre and extraction solid angle used for the X-ray data, shown in Fig.~\ref{fig:layout},
to derive a pressure profile directly comparable to the X-ray one. Errors are computed as the scatter of
radial profiles centered on random positions around the cluster. 
Figure~\ref{fig:surf_bright} shows the SZ surface brightness profile. At $\sim17$ arcmin (about $r_{200}$, or 1000 arcsec), the cluster signal is well above the background. 

The pressure profiles was derived fitting these data using the modified version of \texttt{PreProFit} (Castagna \& Andreon 2019)
used in Andreon et al. (2021).  \texttt{PreProFit} is a Bayesian forward-modeling projection code that fit the
SZ data accounting for the PSF and a possible non-zero
background (named pedestal) level. To 
allow the shape of the pressure profile to vary almost arbitrarily, and, at the same time, be continuous and doubly-differentiable, we modelled the pressure profile as a cubic spline in log-log space with knots at radii of $r=100,500,750$ and $1100$ arcsec (about $180,900,1300$ and $2000$ Mpc). Our model has 5 variables: the pressures at these four points, $P_0,P_1,P_2,$ and $P_3$, and the pedestal value of the SZ surface brightness. By defining
the spline in log quantities (log pressure vs log radius), we naturally exclude non-physical (negative) values
of pressure and radius and we can approximate a large variety of profiles.

We find $P_i=(8.1\pm6.1) \ 10^{-3}, (1.1\pm0.3) \ 10^{-3}, (3.7\pm1.0) \ 10^{-4}, (6.7\pm3.3) \ 10^{-5}$ keV cm$^{-3}$, 
and a pedestal level of $(3\pm2) \ 10^{-7}$  with some covariance, fully accounted for in our analysis. 
We found a $\chi^2$ of $1.1$ for 3
degrees of freedom. Figure~\ref{fig:surf_bright} shows the best-fit model, with the 68\% uncertainty, on top of the observed data. 
The derived pressure profile, with the 68\% uncertainty, is shown and discussed in
Sect.~2.3

\subsection{X-ray data}

\begin{figure}
\centerline{
\includegraphics[width=8.5truecm]{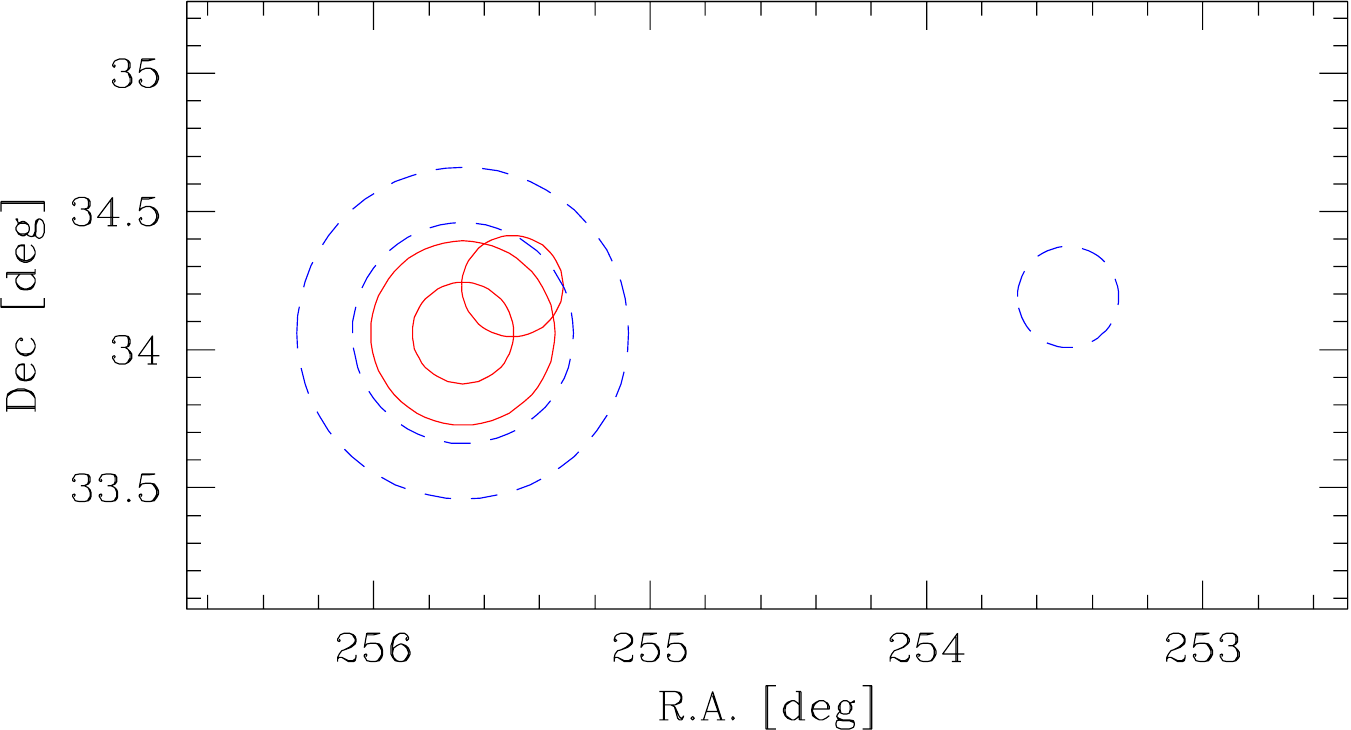}}
\caption[h]{Field layout. Swift and ROSAT observations are indicated by small and large circles, respectively. Regions used for background estimation are indicated with dashed blue circles, those used for cluster measurements by solid red circles. 27 more XRT pointings, not shown in the figure, are used to measure the shape of the background radial profile.
}
\label{fig:layout}
\end{figure}

Swift observed A2244 between November 2011 and September 2012, for a total of 240 ks. Observations have been 
split in two aim points, one 20 ks long (after flare filtering) and close to the cluster center plus a second
212 ks long pointing 15' NW (small circles in Fig.~\ref{fig:layout}). As mentioned,
the determination of the thermodynamic profiles at radii where the background dominates is systematic-limited, and
for this reason an accurate and precise background is needed.
Since the field of view of the cluster observations are entirely filled with the cluster emission,
we considered some external empty fields to estimate the background.
We selected the 27 deep follow-up GRB observations performed in the period 2008/2014 (no more than 3 years
from the cluster observations) at high galactic latitude (abs(lat)$>$ 20 deg) and far away from the Fermi bubble
(40 $<$gal. long  $<$ 320). After the exclusion of the first segments, where the GRB afterglow is bright,
the exposure times are between 90 and 250 ks. As shown in Moretti et al. (2009, 2011, 2012), this data-set can be considered a fair representation of the different components of the Swift-XRT background.
Furthermore, as explained below,  to estimate the Galactic foreground, we  made use of  a
shorter exposure (GRB150213B, 24 ks) which is at only 1.8 deg (about 12 Mpc)
away from A2244 (see Fig.\ref{fig:layout}).

\begin{figure}
\centerline{
\includegraphics[width=8.5truecm]{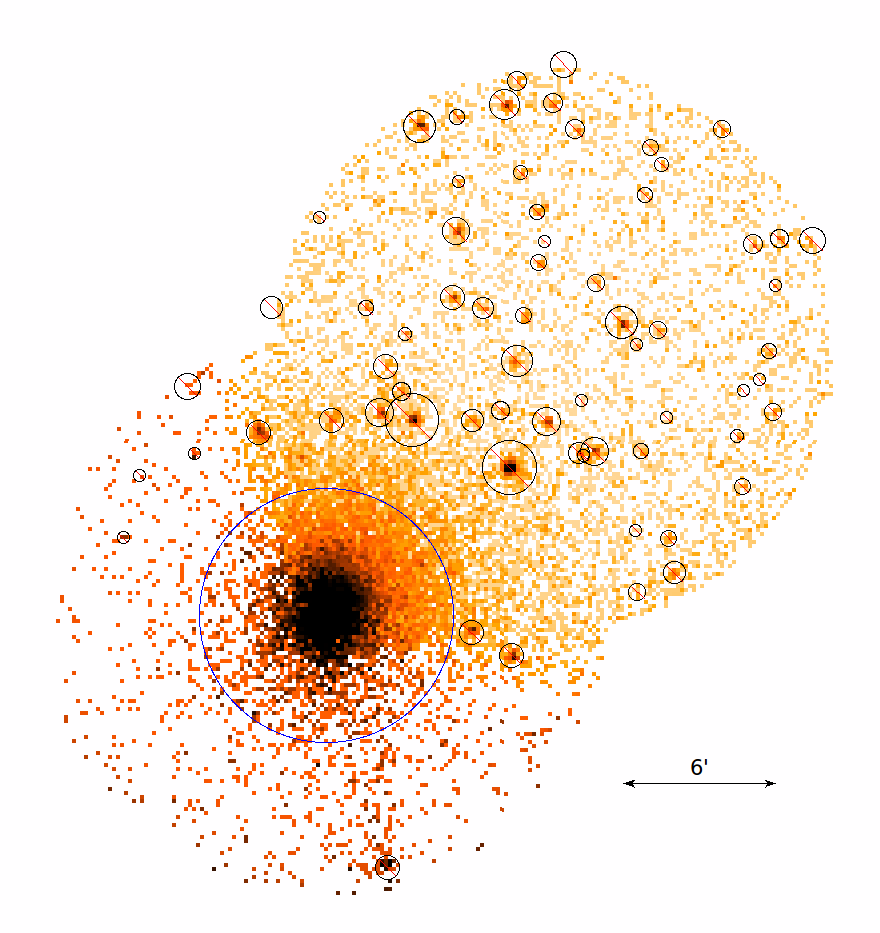}}
\caption[h]{A2244 in the 0.8-1.5 keV band exposure-corrected rebinned to emphasize low surface brightness features. The small black circles indicate flagged sources, whereas the large blue circle indicates the radius outside which we use the NW pointing for the radial profile computation. Emission can be traced even by eye up to 12 arcmin from the cluster center (i.e. to about the center of the NW pointing). The  pointing centered on the cluster (bottom, in the figure) looks noisier because of its reduced exposure time. North is up and East is to the left.}
\label{fig:ima}
\end{figure}

Swift XRT data are reduced following the standard procedures (e.g., Moretti et al. 2009), as
updated in Andreon et al. (2019).
We briefly review the data reduction here. The data are first filtered to
remove periods affected by bright Earth light flares or by increased CCD temperature.
Subsequently, point sources are detected via a wavelet detection algorithm and are masked. Events (photons)
are partitioned into eight bands ([0.8-1.2], [1.2-1.5], [1.5-2.0], [2.0-3.0], [3.0-4.5], [4.5-5.5], [5.5-6.3], and [6.3-7.0] keV). 
We computed
energy-dependent exposure maps to calculate the effective exposure time, accounting for
dithering, vignetting, CCD defects, gaps, and flagged pixels. A cluster image in the [0.5-2] keV band, for display purposes only, is shown in Fig.~\ref{fig:ima}. 

At the time of the A2244 (and control field) observations, the XRT radio-active calibrating sources are still active
affecting the CCD external columns at energies $>5.5$ keV. For this reason,
below 5.5 keV we use the full XRT field of view, whereas at higher energies  
we discarded the first and last 90 columns (unlike the more recent observations considered
in Andreon et al. 2019).
Our current choice saves all the integration time in the outer parts of the field of view
at energies below 5.5 keV. 

We then measured counts and
effective exposure time in the eight energy bands in circular annuli with increasing width 
to counterbalance the decreasing intensity of the cluster.  
The cluster centre is iteratively
computed as the centroid of X-ray emission within the inner 25 kpc radius. The minimal width 
is taken to be 10 arcsec, comparable to the XRT
PSF. We only considered radii where vignetting is less than 50\%.
We combine the information of the two exposures by merging the cluster radial profiles derived in the two pointings using a separation radius of 5 arcmin. The adopted radius is
a trade off: the central pointing features lower S/N at large radii than the
NW pointing because of its reduced exposure time, but, at the same time, at small radii the solid angle sampled 
with the NW pointing is small and effective exposure time is strongly changing.

\begin{figure}
\centerline{\includegraphics[trim=0 0 30 40,clip,width=9truecm]{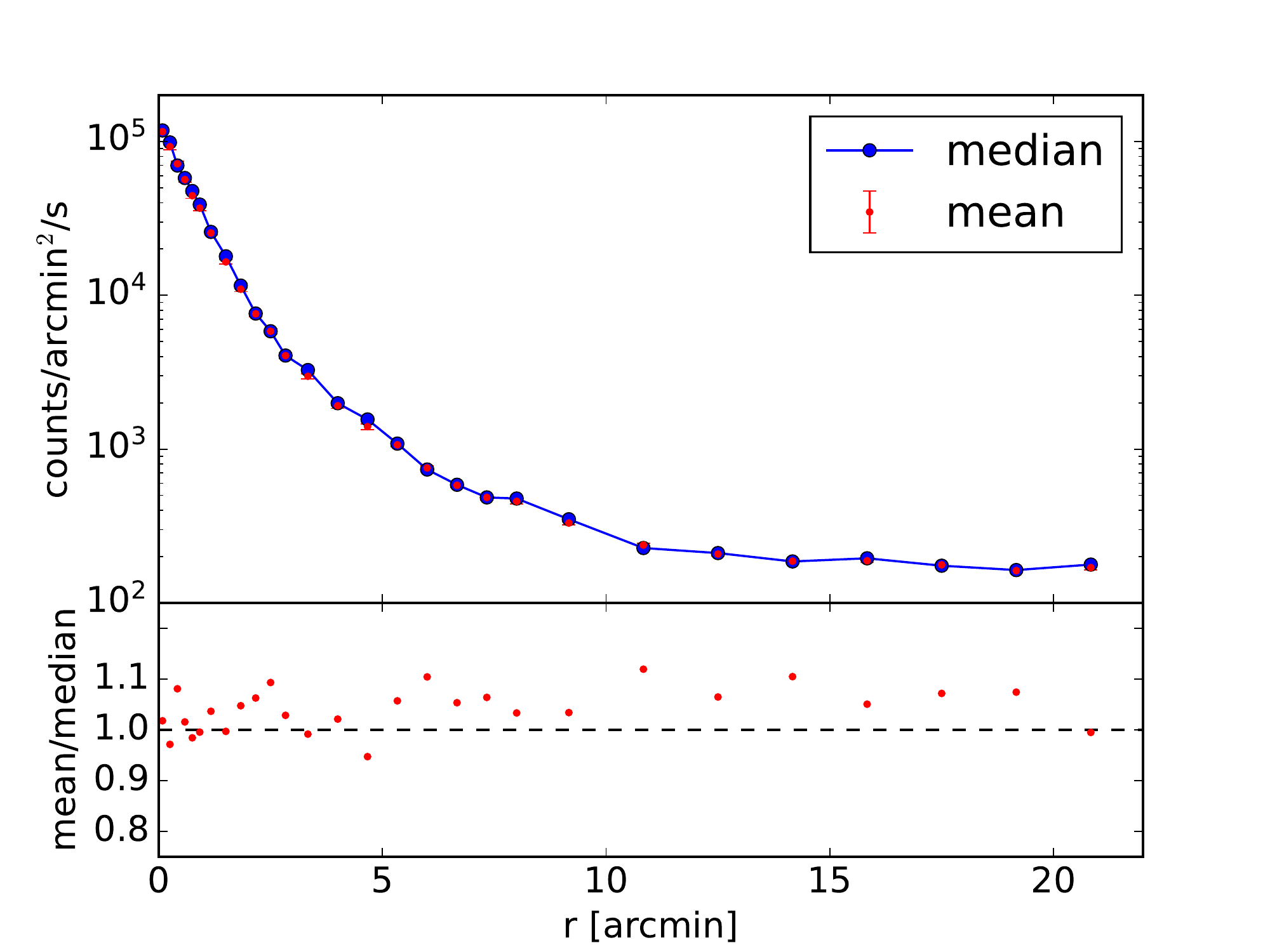}}
\caption[h]{A2244 clumping estimate as derived from the ratio of the azimuthal mean/median profiles (bottom panel). The upper panel shows the two profiles. There is, at most, a 5\% change in the mean/median value in the studied radial range, indicating a low
level of clumping.}
\label{fig:clump_ne}
\end{figure}

If the ICM is clumped, X-ray based quantities, in particular electron density, are biased (e.g. Eckert et al. 2015).
Clumping can be estimated from the ratio between the azimuthal mean and median surface brightness profiles after
the image is adaptively pixellized with pixels with $\geq 20$ counts (Eckert et al. 2015).
Following this path, 
we found in the [0.5-2] keV band a ratio of 
$1.016\pm0.031$ (mean and observed scatter) with $\geq 40$ counts/pixel binning and of 
$1.04\pm0.04$ with $\geq 20$ counts/pixel binning (see Fig.~\ref{fig:clump_ne} for the latter), both without evidence of a radial trend. Given the per cent difference between mean and median, clumping effects are negligible. 
Given the closeness of mean and median profiles for A2244 and that the line of sight projection
needs the mean (times the area) to infer the contribution of outer shells projected at the considered radius (see Andreon et al. 2019 for details), our subsequent analysis uses the mean.

As mentioned, to measure the outer region of the cluster, a systematic-free estimate of the background is paramount.
The Swift XRT  background consists of different components with relative contributions
varying both in energy and within the field of view (Moretti et al. 2009).
Some of them, like the Galactic foreground and cosmic X-ray background 
are vignetted; the others, the NXB and stray-light have different spatial distributions
(Moretti et al. 2009).
Moreover the Galactic foreground is highly variable (Snowden et al 1997).
In each of the eight energy bands, we estimated the shape
of the background profile from the 27 empty fields.
Then, the profiles are 
renormalized to match the (noisy) radial profiles
derived from the nearby background pointing.
To precisely reproduce the spatial-spectral dependence of the background, we adopted the
same geometry used for the cluster analysis: we extracted both profiles, adopting the
same centers in image pixels used for the cluster, and merged the two profiles at the same
separation radius of 5 arcmin.

As mentioned, the vignetted and unvignetted components of the background show differences in their relative intensity, for example depending on the sky coordinates or solar activity. This complicates the spectral analyses commonly performed in the literature because of no single (valid over all energies) 
renormalization may match the spectral shape of background exposures taken with different proportions of the vignetted and unvignetted components. 
Our determination of the radial shape of the background profile is instead unaffected by the varying proportions 
for two reasons: first, the choice of using narrow energy bands makes most of the bands dominated by either the vignetted or unvignetted background components. In such condition, the computation of the shape of the average radial profile, when each pointing is scaled by the mean intensity in the considered narrow energy band, is clearly robust against to intensity variation from pointing to pointing. Second, the off-axis
dependence of the vignetted background components is washed out by our choice of putting the cluster at the boundary of the pointing (the radial profile is an average over a corona including several off-axis angles) or of only considering the central part
of the pointing less affected by vignetting (at small radii). Indeed,
a posteriori, the radial profile turned out to be 
nearly constant in all bands.
Because we basically
work at fixed energy (in narrow energy bands, to be rigorous) and we reduced the impact
of vignetting by effectively marginalizing over off-axis angles,
possible variations of the proportion of vignetted and unvignetted background components commonly complicating spectral analyses are properly handled in our approach.

\begin{figure}
\centerline{
\includegraphics[width=8.5truecm]{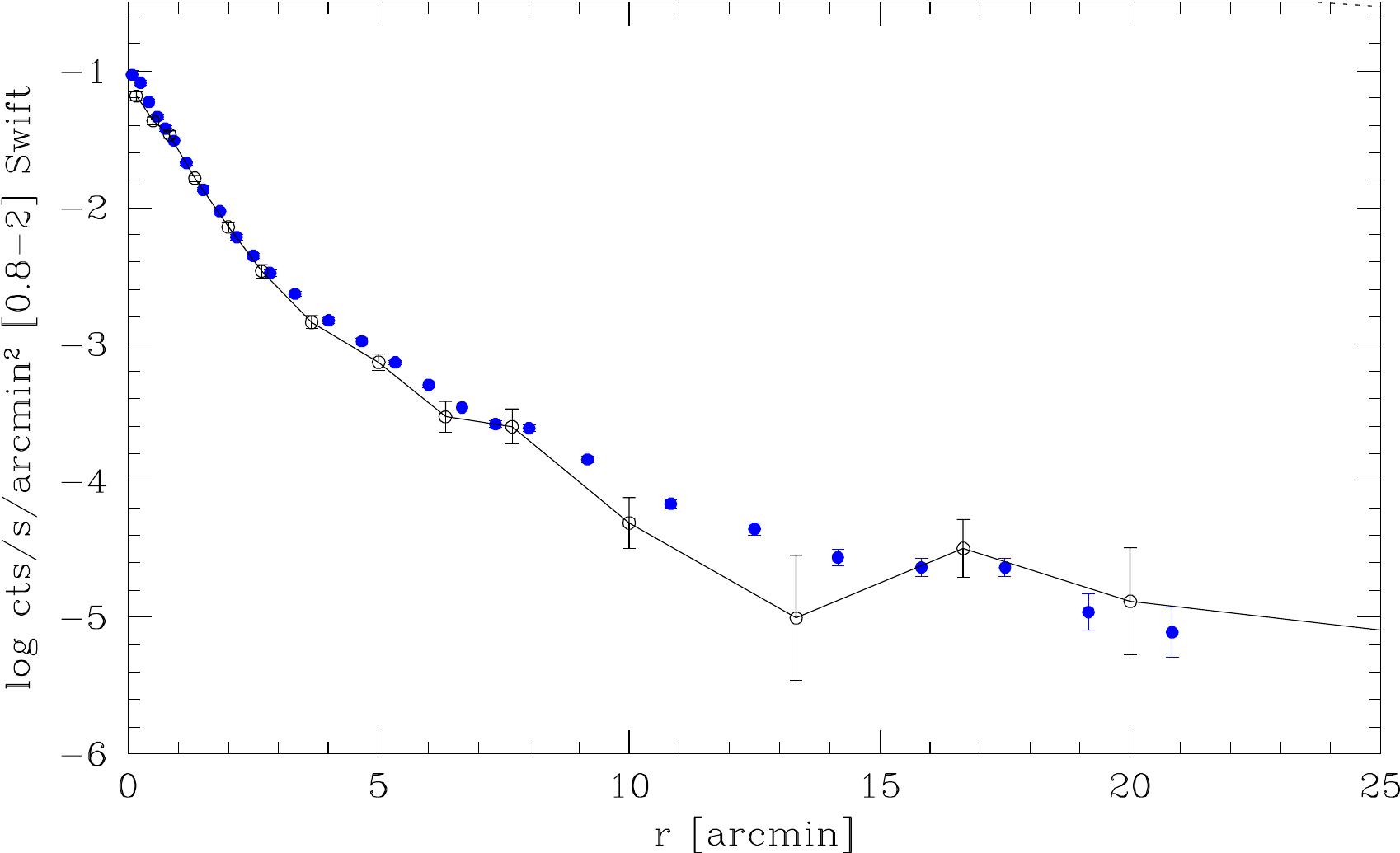}}
\caption[h]{A2244 profile in the 0.8-2 band, as determined by using ROSAT (open points) and XRT (filled blue points with barely visible errors at most radii). The cluster extends well beyond
15', and both XRT and ROSAT show significant and agreeing emission in this soft band (with ROSAT using a closer and simultaneous background). The agreement of the two profiles
indicates a correct estimate of the Swift background
level in the soft bands.
}
\label{fig:rosat}
\end{figure}

\begin{figure*}
\includegraphics[trim=0 0 0 30,clip,width=15truecm]{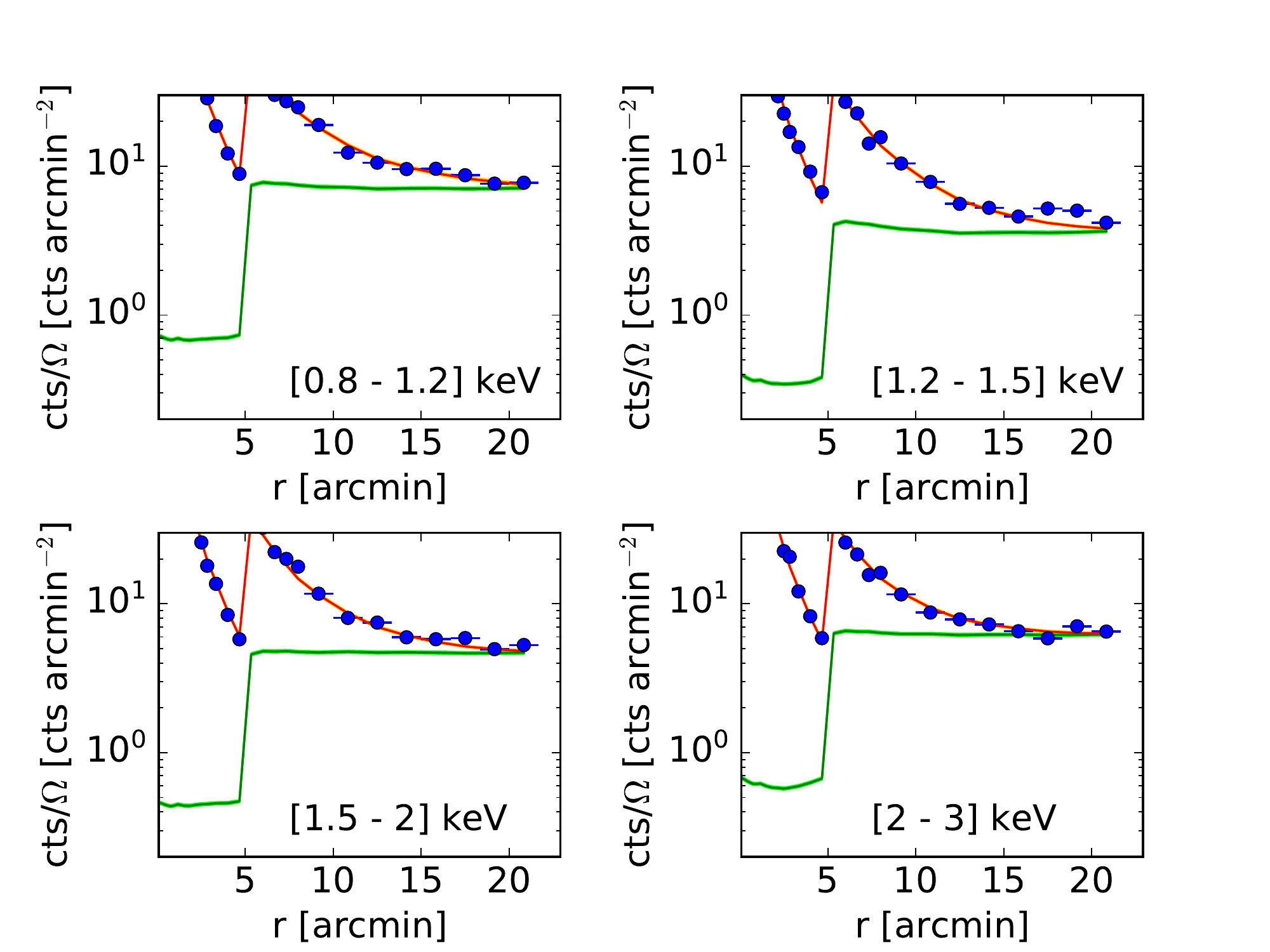}
\includegraphics[trim=0 0 0 30,clip,width=15truecm]{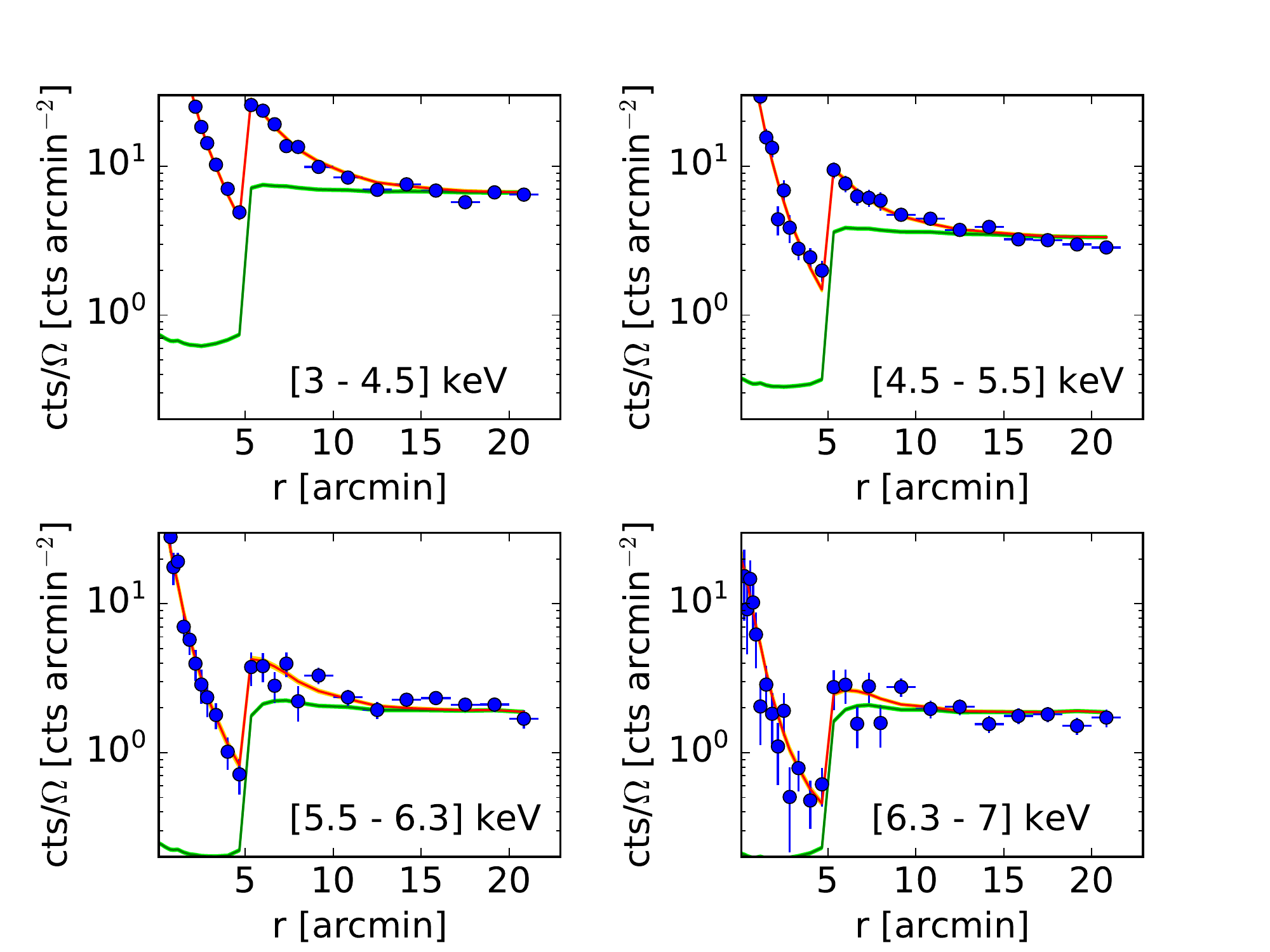}
\caption[h]{Measured brightness in the cluster (points with error bars) and background (green line with
barely visible lime shading indicating 68\% errors) lines of sight before correction for exposure time. 
The red line with barely visible yellow shading is the cluster profile and error fitted to the data. The discontinuity at
5' is caused by the abrupt change of exposure time there (change of pointing). The plot zooms on the 
interesting low-count regime (at smaller radii there are points outside the plotted ordinate range). 
While at $E<2$ keV there is
a clear excess above the background beyond 15', at larger energies the excess disappears. This is the observable used in the fit to infer a low temperature at large radii.
}
\label{fig:radprof}
\end{figure*}

The estimated low energy background can be independently checked using pointed
ROSAT PSPC data. Fig.~\ref{fig:rosat} shows the 0.8-2 keV cluster profile as derived from XRT, and the one derived from the 0.4-2.4 keV 3.0 ks Rosat observation. 
The two considered energy bands are quite close each other, and even more when considering that C-absorption
of the plastic window of the PSPC makes it almost blind in the 0.4-0.5 energy band,
and the mirror efficiency does the same in the energy range 2.0 to 2.4 keV, making  the band
conversion to be temperature independent. The
radial profiles are independently background-subtracted: XRT data as described above, ROSAT data using the annulus
indicated in Fig.~\ref{fig:layout}.
Thanks to the larger PSPC field of view, the ROSAT background is measured from the same pointed
observations, i.e. it is simultaneous and closer in the sky than the one used for XRT (see Fig.~\ref{fig:layout}).
ROSAT countrates are derived the NW quadrant, to approximately match the
XRT field of view, and converted into XRT ones using XSPEC. The two profiles, independently background-subtracted, agree well each other, in particularly at $r\sim r_{200}\sim17$ arcmin,
confirming the appropriateness of our estimate of the low-energy XRT background. 
The larger error bar of the ROSAT profile at $r\sim 13$ arcmin is due to the reduced sensitivity at these radii because of the support of the plastic window of the PSPC.

At higher ($kT>2$ keV) energy, the XRT background is spatially uniform inside the field of view
and stable in time and across fields (Moretti et al. 2009, 2011), 
and indeed the observed scatter across the 27 background fields is about 7-8\%. Although 
we have no high energy photons from independent telescopes to check the XRT background at higher energies, we will indirectly do it in Sec.~2.3 by comparing the pressure
profile derived from X-ray and SZ data: if the high energy background estimate is incorrect, a wrong X-ray temperature
would be measured and, as a consequence, a difference should appear between pressure (density times temperature) derived in X-ray
and from SZ data.

As a final precaution, our analysis allows the background in the cluster line
of sight to differ from the one outside the cluster by a factor to be
determined with the data.

To summarize, we have two (one cluster and one background) X-ray data cubes with 8 energy bins
and 30 radial bins, that we plot as radial profiles in different bands, 
shown in Fig.~\ref{fig:radprof} (points; the step is due to the
changing $\sim10\times$ exposure time at the profile junction). 
Each cube may also be alternatively seen as
eight-points spectra taken at 30 different clustercentric distances. 

There are 18600 net photons within 18 arcmin from the cluster center in
the [0.8-7] keV band.

\subsection{Analysis}

A change in the cluster temperature shows up in X-ray data space (i.e. in Fig.~\ref{fig:radprof}) 
as a non-constant ratio between net count-rates measured in different bands. Close inspection of 
Fig.~\ref{fig:radprof} suggests this could be the case:
for $E<2.0$ the cluster radial
profile shows an excess of counts over the background at large radii (say
17 arcmin, i.e. $r_{200}$), absent at higher energies. This means that at these distances the gas is colder than
at smaller distances from the cluster center. This needs, however, to be quantitatively assessed.

\begin{figure}
\centerline{
\includegraphics[width=8.5truecm]{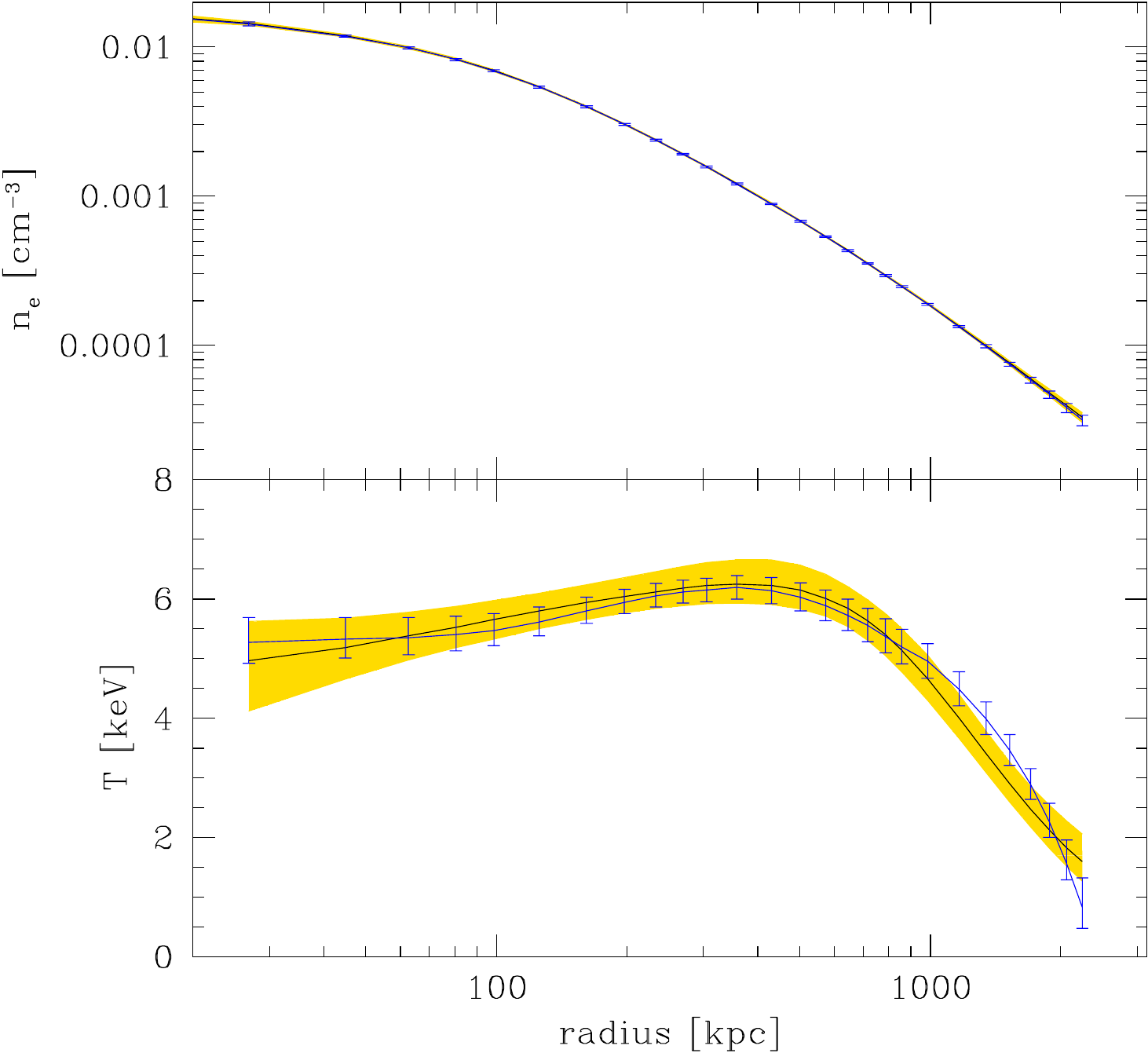}}
\caption[h]{A2244 deprojected electron density and temperature profiles. The solid line with shading
shows our descriptive fit, while the solid line with error bar is our physical fit. 
In both cases we plot the mean model and 68\% errors. These thermodynamic profiles are,
by large, model-independent.}
\label{fig:thermo_profs_ne_T}
\end{figure}

Quantitatively, 
thermodynamic profiles were derived using \texttt{MBProj2} (Sanders et al. 2018), a Bayesian
forward-modeling projection code that fits the data cube accounting for the background. 
Similarly to other approaches, \texttt{MBProj2} assumes  
spherical geometry, lack of clumping, and, when requested, hydrostatic equilibrium. \texttt{MBProj2} overcomes the limitations commonly found in literature, such as 
the use of arbitrary regularization kernels to deal with noise, ignoring
temperature gradients when deriving the electron density profile, 
or ignoring the spectral variation of
the background.

To understand the impact of modeling choices on
the results we fit the data with two models. First, we follow the descriptive approach of Vikhlinin et al. (2006): we model the temperature
and electron density profiles with flexible functions constrained by the data. This is a 
fit aimed at simply describing the observations, with models introduced to impose 
regularity and smoothness. 
In particular, we use a modified single-$\beta$ profile for the electron
density:
\begin{equation}
  n_\mathrm{e}^2 = n_0^2
  \frac{(r/r_\mathrm{c})^{-\alpha}}{(1+r^2 / r_\mathrm{c}^2)^{3\beta-\alpha/2}}
  \frac{1}{(1 + r^\gamma / r_\mathrm{s}^\gamma)^{\epsilon / \gamma}}.
\label{eq:dens}  
\end{equation}
and 
similarly to Vikhlinin et al. (2006), the temperature profile is given by 
\begin{equation}
\rm{T} = \textrm{T}_0\frac{((r/r'_c)^{a_{cool}}+(\textrm{T}_{min}/\textrm{T}_0))}
         {(1+(r/r'_c)^{a_{cool}})}\frac{(r/r_t)^{-a}}{(1+(r/r_t)^b)^{c/b}}.
\end{equation}

The other
thermodynamic profiles are derived from the ideal gas law. 
Following Vikhinin et al. (2006), we fix $\gamma=3$ and
following McDonald  et al. (2014), we
fix the inner slope to $a=0$ and the shape parameter of the inner region to $a_{cool}=2$.
We adopt weak priors for the remaining parameters. 

In the second modeling, we 
adopt a Navarro et al. (1997; hereafter NFW) profile for the dark matter
and assume hydrostatic equilibrium. In this case, pressure is derived from mass
and electron density profiles via the hydrostatic equilibrium equation 
and the other thermodynamic profiles follow from
the ideal gas law. There is no direct temperature profile modeling in this approach,
only the indirect one coming from the hydrostatic equilibrium assumption.

In all our fits, metallicity is a free parameter, 
absorption was fixed at the Galactic $N_H$ value in the direction of the cluster
from Kalberla et al. (2005),
and the results are marginalized over a further background scaling parameter to account
for systematics (differences in background level between the cluster and
control fields, see Sanders et al. 2018 for details), taken to have a prior centered on one with 10\% sigma, close to the observed 
background scatter across fields.
The model is integrated on the same energy 
and radial bins as the observations, so that the results do not depend on the
binning choice.

Figure~\ref{fig:radprof} illustrates how well the descriptive model (red line with 68\% 
uncertainty shaded) fit the data. The plot for the physical fit is indistinguishable.
Both descriptive and physical model fit well in a $\chi^2$ sense
($\chi^2_\nu\sim1.2$). 
The backscale parameter, that
measure the amplitude of a potential (multiplicative) offset between the X-ray backgrounds
in lines of sight of the cluster and empty fields, is extremely well determined and very
close to one,  $0.97\pm 0.02$. 
The gas metallicity is  $Z_{Fe}=0.34\pm0.07$ times the Solar value given in Anders \& Grevesse (1989).

\begin{figure}
\centerline{\includegraphics[width=8.5truecm]{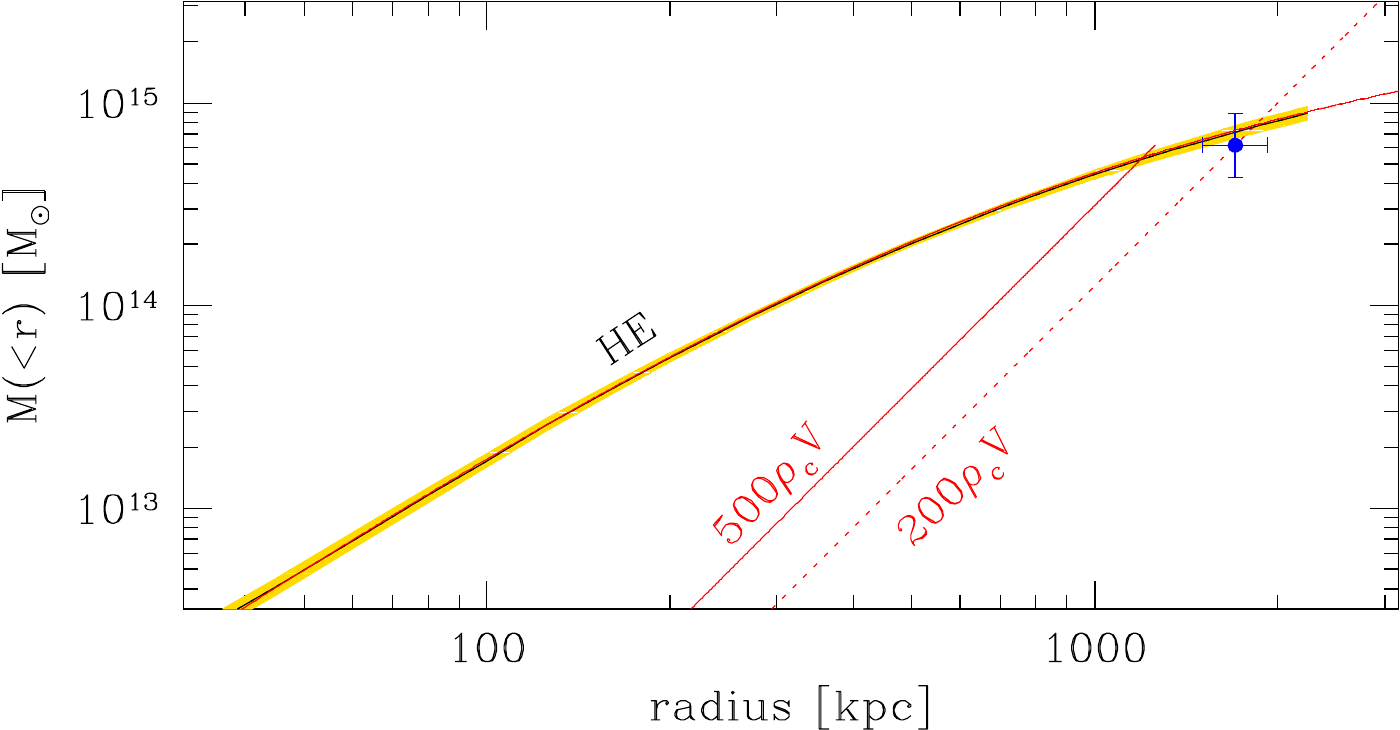}}
\caption[h]{A2244 total mass profile with 68\% uncertainties (solid curve and shading). The red line indistinguishable from the 
mean profile is a NFW with $c_{200}=4$. The point with error bars indicates the caustic mass estimate by Rines \& Diaferio (2006). Hydrostatic and caustic mass, each one affected by different systematics,
agree well which each other.}
\label{fig:mass}
\end{figure}

The electron density profile, which is proportional to the deprojected surface brightness 
profile in soft bands (upper panel of Fig.~\ref{fig:thermo_profs_ne_T})
is robustly determined and independent of the assumed model and it can be traced to 
$\gtrsim 2$ Mpc at least ($\sim r_{200}$) 
(Fig.~\ref{fig:radprof}).  The $n_e$ slope near $r_{200}$
is $\sim -2.2$ in the log-log plane, close to numerical simulation expectations (Roncarelli et al. 2006).

The X-ray temperature profile (bottom panel of Fig.~\ref{fig:thermo_profs_ne_T}) is robustly measured too, 
the sharp decline at large
clustercentric distances comes from the cluster fading faster with radius in higher energy bands
than in lower ones, as already mentioned. The temperature
profile derived with the physical fit, i.e. without a direct modeling of the temperature profile, is fully
consistent with the one derived with
our descriptive fit. 
The independence
of the derived of the temperature profile shape from its modeling indicates that the derived profile
is driven by the data and not
forced by the adopted mathematical expression for the temperature model.
The nearly isothermal profile at $r<500$ kpc with about 5.5 keV temperature 
closely agrees with the Chandra estimate of $5.5\pm0.5$ keV (Donahue et al. 2005).

Fig.~\ref{fig:mass} shows the total (i.e. dark matter plus gas) mass profile derived from the X-ray data assuming hydrostatic equilibrium.
We derived $\log r_{200}=3.26\pm0.02$ and $\log r_{500} =3.06\pm0.01$ kpc.
The red line, indistinguishable from the best fit model (in black), shows a NFW with $\log M_{500,HE}/M_\odot=14.68\pm0.04$, $\log M_{200,HE}/M_\odot=14.89\pm0.02$  
and $c_{200}=4.0$.
Fig.~\ref{fig:mass} shows that the found hydrostatic $M_{200,HE}$ perfectly agrees with the mass
derived with the caustic technique: $\log M_{200,HE}/M_\odot=14.79\pm0.17$ (Rines \& Diaferio 2006). The latter is
based on galaxy dynamics, free from assumptions about the cluster state and about gas-related
properties, such as non-thermal pressure support and gas clumping. The remarkable agreeement between
the two mass estimates indicates a low value of hydrostatic mass bias, and a low level of 
non-thermal pressure support and clumping (or very good luck).

\begin{figure}
\centerline{\includegraphics[trim=0 0 0 0,clip,width=8.5truecm]{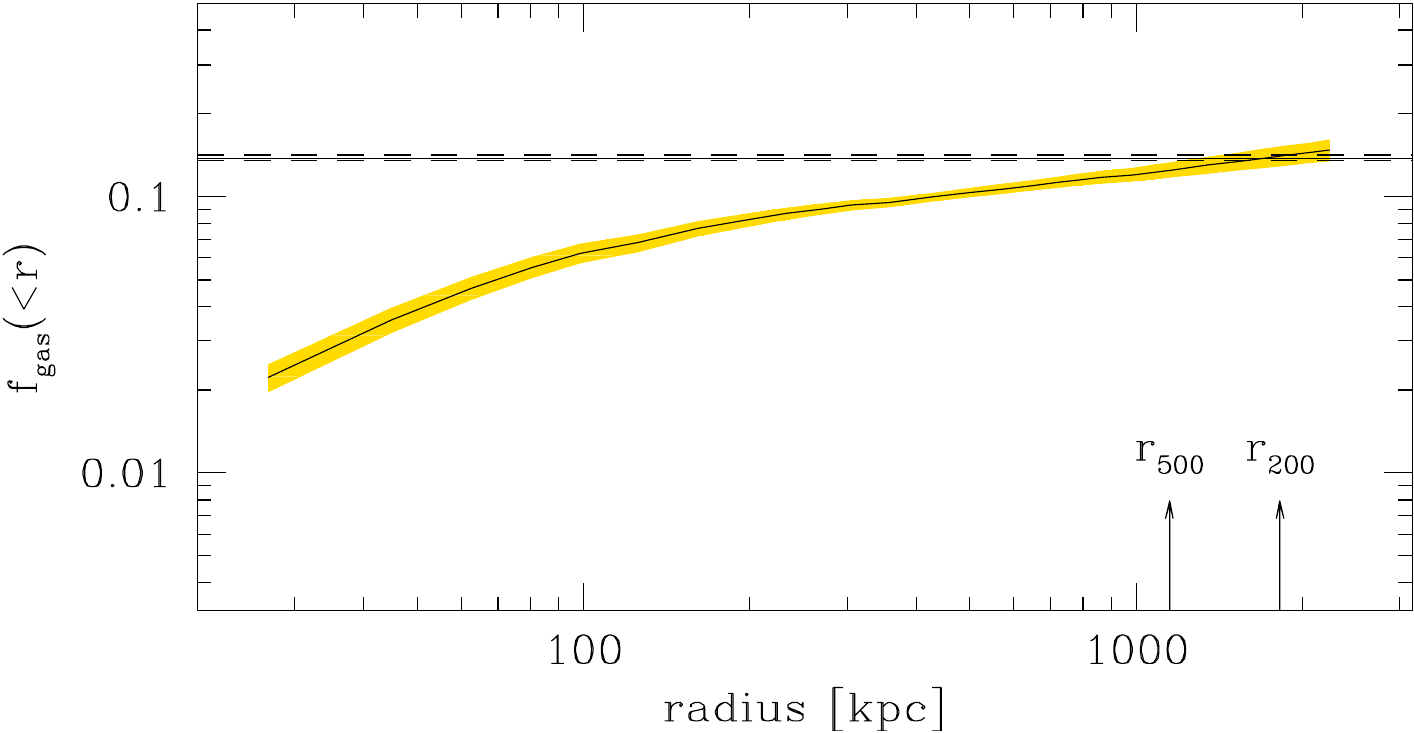}}
\caption[h]{A2244 cumulative gas fraction profile with 68\% uncertainties (solid curve and shading). The horizontal lines indicate
the Universe baryon fraction and uncertainty from Planck collab. (2015) minus
the baryon fraction in stars as measured in Andreon (2010). Unlike determinations of some other clusters with gas density
profiles exceeding the Universe baryon fraction,
the total baryon fraction stays close to the Universal one at
$r_{200}$ confirming once more the low level of clumping.}
\label{fig:fgas}
\end{figure}

Fig.~\ref{fig:fgas} shows the gas mass fraction (solid line with shaded 68\% errors). We also report the expected
gas mass fraction, given by the Universe baryon fraction (Planck Collaboration 2015) minus the A2244 stellar mass
fraction ($0.018\pm0.002$, from Andreon 2010). 
The gas mass fraction
converges to the value expected from the Universe baryon fraction at $r_{500}<r<r_{200}$.
From it, we derive a $<2$\% clumping at $r_{200}$ under the assumption that the total baryon
fraction equals the Universe value, in agreement with our earlier estimations of clumping.

\begin{figure}
\centerline{
\includegraphics[width=8.5truecm]{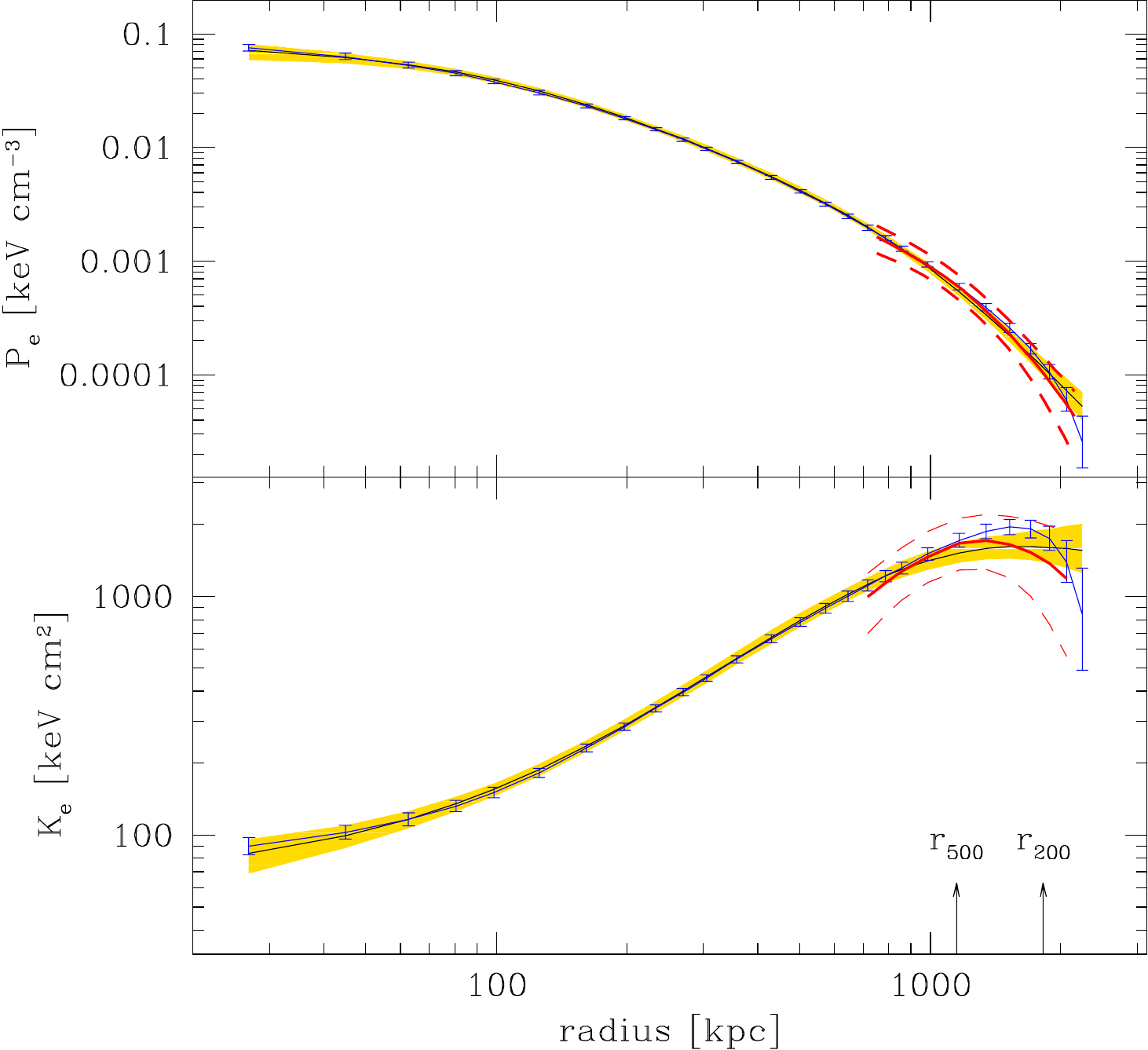}}
\caption[h]{A2244 pressure and entropy profiles. 
 The solid line with shading
shows our descriptive  fit, while the solid line with error bar is our physical fit.
In the top panel, we also plot the SZ-based pressure profile and its 68\% error (solid line and dashed corridor, respectively).
In the bottom panel,
the red solid line combines the SZ pressure profile to the X-ray $n_e$ profile, while the dashed lines show its 68\% uncertainty. These thermodynamic profiles are model- and observable-independent. 
}
\label{fig:thermo_profs_P_K}
\end{figure}

The X-ray pressure profile (upper panel of Fig.~\ref{fig:thermo_profs_P_K})
turns out to also be independent of the fitted model, and, furthermore, to agree with the independent
derivation based on SZ data (red solid line with 68\% bound shown as dashed corridor). 
Since pressure
is the product of electron density and temperature, the good agreement between X-ray and SZ-based pressure profiles
confirms the observed
decrease of temperature at radii where we cannot test the XRT high-energy profile with X-ray data
from other telescopes. Furthermore, since SZ pressure is $n_e$-weighted while X-ray pressure is $n^2_e$-weighted,
the agreement between the two pressure determinations indicates than clumping $\langle n^2_e\rangle /\langle n_e\rangle^2$
is close to one, in agreement with other evidences already presented.

Finally, the bottom panel of Fig.~\ref{fig:thermo_profs_P_K} shows
the entropy profile using the two X-ray modelings (solid red line with yellow shading and points with blue error bars). It also
shows the entropy profile derived combining the X-ray $n_e$ profile
with the SZ pressure profile, i.e., which replaces X-ray spectral information with SZ data (solid line with dashed corridor). 
The two X-ray data modelings give pretty similar entropy profiles and both
show a clear non-linear entropy profile (in a log-log scale) at larger radii. This bending is robust because the electron density
profile is well determined even at $r_{200}$ since at this radius and at low energies the cluster signal is large in absolute terms and compared to the low XRT background, see Fig.~\ref{fig:radprof}, whereas 
the range of possible temperatures is strongly reduced because of the negligible cluster 
signal at $kT>3$ at those radii.
The mixed SZ-X-ray derivation, which has a different sensitivity to clumping, also
shows a similar bending of the entropy profile, confirming that the X-ray derivation is robustly determined.
At small radii, the elevated value of entropy in the cluster core agrees with Donahue et al. (2005). 
At large radii, the A2244 entropy profile shows a bending similar to the one found
in several clusters using the low X-ray background Suzaku telescope (Walker et al.
2019) and to the A2319 profile derived from XMM-Newton (Ghirardini et al. 2018), although 
A2244 has not the complex morphology of A2319 neither its larger
than universal gas mass fraction value.

\begin{figure}
\centerline{\includegraphics[width=8.5truecm]{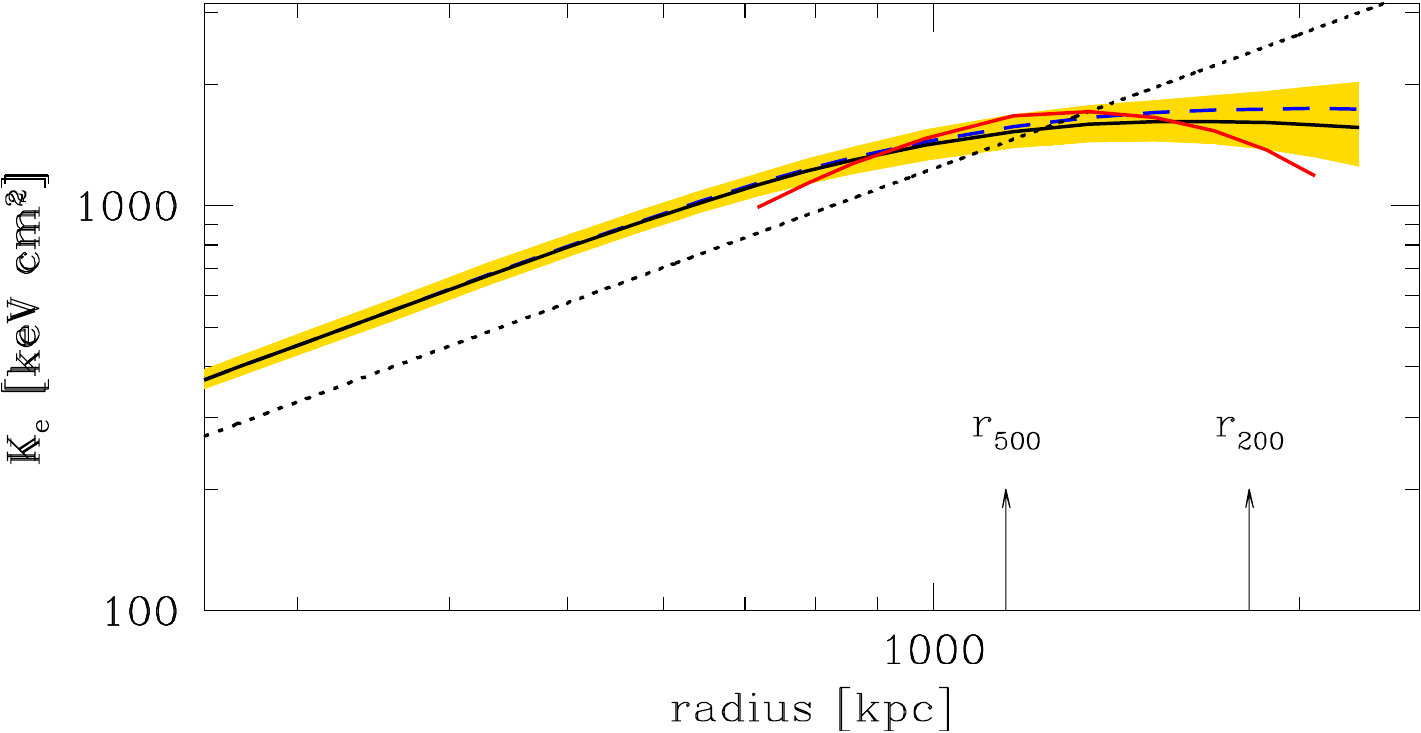}}
\caption[h]{Zoom-in on the region showing a flattening in the entropy profile. The solid black
line with shading shows our descriptive fit, while the blue dashed line shows the entropy profile corrected for electron-ion non-equilibration according to Avestruz et al. (2015) simulations.
The dashed line is the Voit et al. (2005) fit to non-radiative simulations, as adapted by Pratt et al. (2010). A $10$ per cent correction for deviation from equilibration cannot restore a power law entropy profile, nor does a few per cent (at $r_{200}$) correction for electron density clumping. The red line shows
the entropy profile obtained combining SZ and X-ray derivations confirming that the derived profile is observable-independent.}
\label{fig:K_zoom}
\end{figure}

Fig.\ref{fig:K_zoom} zooms at large radii comparing  the hypothesis-parsimonious descriptive fit  to the  power law expected from the hierarchical clustering  fit to non-radiative simulations (Voit et al. 2005, as adapted by Pratt et al. 2010). There is
an obvious excess within $r_{500}$ common to most clusters (e.g. Ghirardini et al. 2020) and a bending at larger radii. At large radii electrons
could be colder than ions because of the long equilibration timescale (e.g.
Hoshino et al.  2010; Akamatsu et al. 2011). The dashed line shows the entropy profile corrected for ion-electron equilibration according to Avestruz et al. (2015) simulations. As it is clear from this figure, this minor correction, about 10\% at $r_{200}$, is largely insufficient to restore a power law entropy profile. In the same vein, the few per cent (at most) bias in surface brightness due to clumping (Sec.~2.2) affects the
entropy profile at most at the per cent level and cannot restore a power-law entropy profile. The red line shows
the entropy profile obtained combining SZ and X-ray derivations, confirming that the derived profile is observable-independent.

\begin{figure}
\centerline{
\includegraphics[width=8.5truecm]{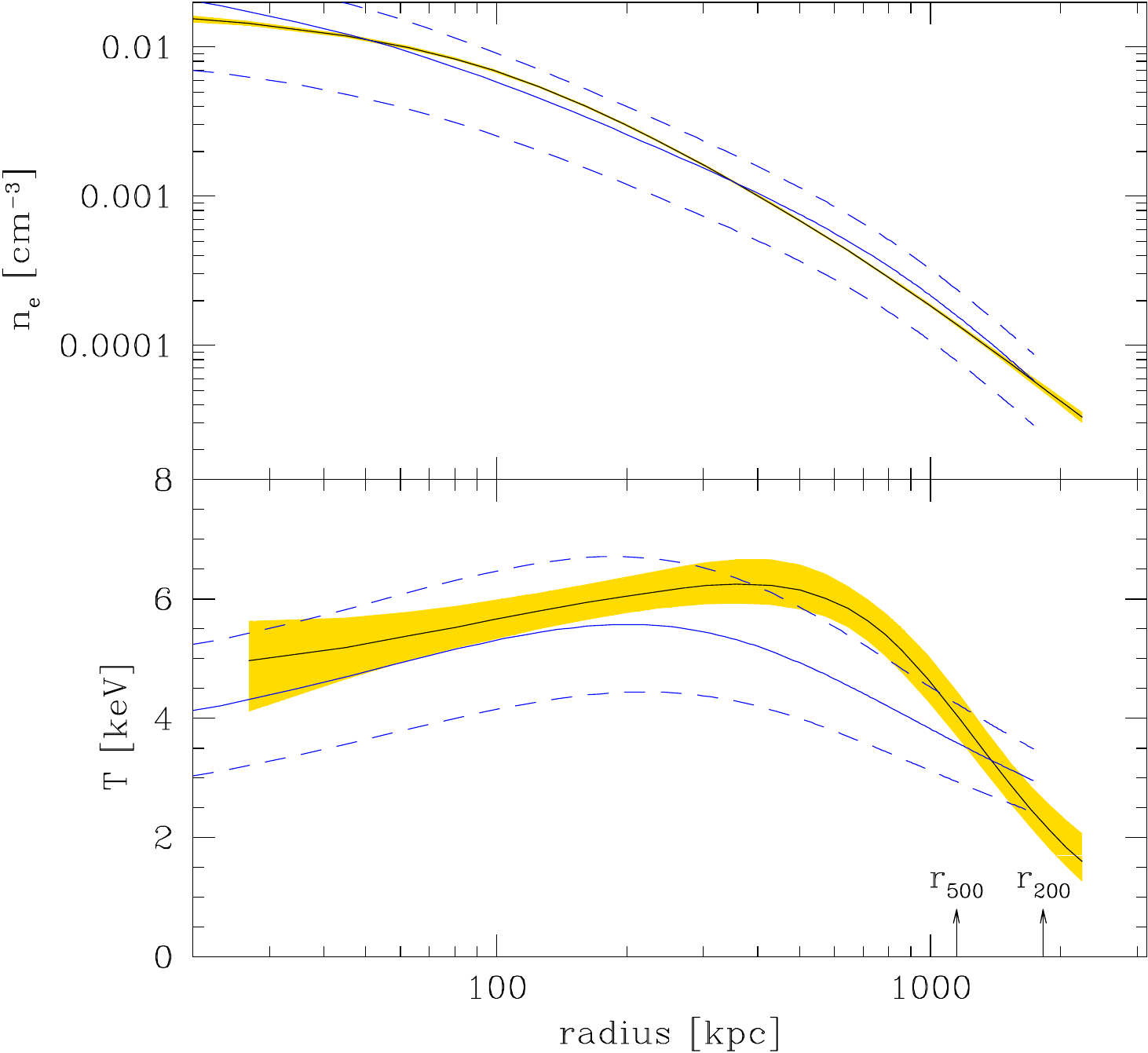}}
\caption[h]{A2244 electron density and temperature profiles from our descriptive fit (as in Fig.~\ref{fig:thermo_profs_ne_T}) with superposed
mean (solid line) and $\pm2\sigma$ bounds of the X-COP sample. The A2244 temperature profile is
much steeper at large radii than the mean X-COP profile (dashed corridor). }
\label{fig:ne_T_XCOP}
\end{figure}

In order to determine which one between density and temperature contribute most to the entropy flattening, Fig.\ref{fig:ne_T_XCOP}
compares A2244 profiles to the mean profiles derived by X-COP (Ghirardini et al. 2019). The A2244 electron density profile is pretty close
to the X-COP mean and, in particular, has similar slope at large radii. 
A radially increasingly clumped ICM would flatten it. As mentioned, the azimuthal mean/median test 
shows that clumping in A2244 is modest at most and radially-independent. The agreement of the slope of A2244 profile, based on azimuthal
mean, with X-COP one, based on median, independently confirms the small role of clumping in A2244.
The outer temperature profile is instead steeper in A2244 than in the X-COP mean, but not unusual since Abell 644, one of the 13 X-COP clusters, shows a similar sharp decrease near $r_{200}$. 

Therefore, the bending of the entropy profiles seems to
be robustly derived and we therefore interpret it as due to exhaustion of the mass inflow through the virial boundary, which reduces entropy production, flattens the entropy profile (Lapi et al. 2010; Cavaliere et al. 2011) and induces sharp declining 
temperature profiles.

\section{Conclusions}

In this work
we derived the entropy profile to large radii of the Abell 2244 galaxy cluster both exclusively using X-ray data from the low-background Swift XRT telescope and also combining them with Planck $y$ data. The entropy profile derivation using X-ray only is robust at least to the virial radius because the cluster brightness is large compared to the X-ray background at low energies, temperature is strongly bounded by the lack of cluster X-ray photons at energies $kT>3$ keV, and the XRT  background is low, stable and understood. 
The background at low energies was also independently checked using ROSAT photons confirming the cluster brightness
at the virial radius.

In the observed solid angle, about one quadrant,
the entropy radial profile deviates from a power-law at the virial radius, mainly because of a sharp drop of the cluster temperature. 
This bending of the entropy profile is confirmed when X-ray spectral information is
replaced by the Compton map. 
Clumping and non-thermal pressure support
are insufficient to restore a power-loaw entropy profile because they are bound to be small by: a) the
agreement between mass estimates from different tracers (gas and galaxies), b) the agreement between entropy profile
determinations based on combinations of observables with different sensitivities and systematics, and c) the low value of clumping as
estimated using the azimuthal scatter and the gas fraction. 

Based on numerical simulations, 
ion-electron equilibration is also insufficient to restore a power-law entropy profile. Therefore, the bending of the entropy profiles seems to be robustly derived and witnesses the teoretically-predicted decrease in the inflow through the 
cluster virial boundary. 

\section*{Acknowledgements}
We thank the referee for asking to clarify how we dealt for the background.
We thank Stefano Ettori, Dominique Eckert, and Fabio Gastaldello for useful discussion.

\section*{Data Availability}
All the data analysed in this work are public available at https://heasarc.gsfc.nasa.gov/ (X-ray) or http://pla.esac.esa.int/pla/ (Planck).

\bsp	
\label{lastpage}

\end{document}